\documentclass[prd,twocolumn,showpacs,preprintnumbers,amsmath,amssymb,superscriptaddress,floatfix,nofootinbib]{revtex4-2}
\usepackage{float}
\usepackage{graphicx}
\usepackage{bm}
\usepackage{amsmath}
\usepackage{amsfonts}
\usepackage{amssymb}
\usepackage{color}
\usepackage{multirow}
\usepackage{subfigure}
\usepackage{caption}
\usepackage[colorlinks, citecolor=blue,anchorcolor=red,menucolor=red, linkcolor=red,filecolor=red,runcolor=red,urlcolor=blue,frenchlinks=red]{hyperref}

\begin{document}
\title{Roles of $a_0(980)$ and $a_0(1710)$ in Cabibbo-suppressed process $D^+\to \pi^0\pi^+\eta $}

\author{Xiao-Hui Zhang}
\affiliation{School of Physics, Zhengzhou University, Zhengzhou 450001, China}\vspace{0.5cm}
	\affiliation{Institute of Frontier and Interdisciplinary Science, Shandong University, Qingdao 266237, China}
	
\author{Jing-Yu Zhu}~\email{zhujingyu@zzu.edu.cn}
	\affiliation{School of Physics, Zhengzhou University, Zhengzhou 450001, China}\vspace{0.5cm}

 \author{Li-Juan Liu}~\email{liulijuan@zzu.edu.cn}
	\affiliation{School of Physics, Zhengzhou University, Zhengzhou 450001, China}\vspace{0.5cm}
 
\author{En Wang}~\email{wangen@zzu.edu.cn}
\affiliation{School of Physics, Zhengzhou University, Zhengzhou 450001, China}
	
	\begin{abstract}
   Motivated by the BESIII amplitude analysis of the single Cabibbo-suppressed process $D^+\to \pi^0\pi^+\eta$, we investigate this reaction by taking into account the contributions from the $a_0(980)$, $\rho$, and $a_0(1710)$, where the scalar meson $a_0(980)$ could be dynamically generated from the $S$-wave pseudoscalar meson-pseudoscalar meson interaction within the chiral unitary approach.
   Our theoretical predictions for the $\pi^0\eta$, $\pi^+\eta$, and $\pi^+\pi^0$ invariant mass distributions are in agreement with the BESIII measurements, especially the clear peaks around 1~GeV in the $\pi^0\eta$ and $\pi^+\eta$ invariant mass distributions could be associated with the dynamically generated state $a_0(980)$. Furthermore, we demonstrate that the intermediate $a_0(1710)$ is also necessary to describe the enhancement structure around $1.6\sim 1.7$~GeV in the $\pi^{0/+}\eta$ invariant mass distribution. More precise experimental measurements of this process could provide deeper insights into the nature of the scalar mesons $a_0(980)$ and $a_0(1710)$.

	\end{abstract}
	
	\pacs{}
	\date{\today}
	
	\maketitle
	
	\section{Introduction}\label{sec1}
In the classical quark model, baryons and mesons are respectively described as three-quark and quark-antiquark bound states. However, recent experimental discoveries have unveiled a plethora of exotic hadron states that challenge conventional quark model interpretations. To elucidate their underlying structure, various theoretical frameworks-including compact tetraquarks, hadronic molecular states, and hybrid configurations—have been proposed~\cite{Gell-Mann:1964ewy,Gao:1999ar,Oset:2016lyh,Lu:2024dtb,Baru:2003qq,Yang:2024idy,Wang:2024jyk}.

The nature of light scalar mesons, particularly the $a_0(980)$, $f_0(980)$, and $f_0(500)$ states, remains enigmatic, and continues to pose significant theoretical challenges~\cite{Achasov:2021dvt,Yu:2020vlt,Ahmed:2020kmp}. Notably, the isovector scalar meson $a_0(1710)$ was first observed by the BaBar Collaboration~\cite{BaBar:2021fkz}, with subsequent confirmations by BESIII~\cite{BESIII:2021anf,BESIII:2022npc} and LHCb~\cite{LHCb:2023evz}. The considerable discrepancies in its measured mass have stimulated extensive theoretical investigations~\cite{Wang:2022pin,Zhu:2022wzk,Zhu:2022guw,Dai:2021owu,Oset:2023hyt,Ding:2023eps,Ding:2024lqk,Wang:2023lia,Wang:2023aza}.

Hadronic decays of charmed hadrons provide a unique window into hadron-hadron interactions and CP violation phenomena~\cite{Cheng:2015iom,Li:2025exm,Li:2025gvo,Li:2024rqb,Duan:2024czu,Zhang:2024jby,Lyu:2024qgc,Wang:2022nac,Lyu:2025oow}. These processes serve as critical testing grounds for quantum chromodynamics (QCD) and short-range weak interactions~\cite{Oset:2016lyh,Wang:2017pxm}. Recent advances by the BESIII and Belle Collaborations have yielded abundant high-precision data on multi-body charmed decays, significantly enhancing our capability to investigate light scalar mesons~\cite{BESIII:2023htx,Ge:2022dzi,BESIII:2021aza,Belle:2020fbd}.

In 2019, the BESIII Collaboration observed the decay process $D^+\to \pi^+\pi^0\eta$ using  an integrated luminosity of 2.93~fb$^{-1}$ collected at the center-of-mass energy of 3.773~GeV, and reported the branching fraction $\mathcal{B}=(2.23\pm0.15\pm0.10)\times 10^{-3}$~\cite{BESIII:2019xhl}. Subsequently, the authors of Ref.~\cite{Ikeno:2021kzf} has studied this process by considering the dynamically generated scalar $a_0(980)$ within the chiral unitary approach, and predicted the $\pi^+\eta$ and $\pi^0\eta$ invariant mass distributions.
  Recently, the BESIII Collaboration performed the first amplitude analysis of the decay $D^+\to \pi^0\pi^+\eta$ using a data sample 7.9~fb$^{-1}$ taken with the BESIII detector at the center-of-mass energy of 3.773 GeV~\cite{BESIII:2024tpv}, and pointed out that the $a_0(980)^+$ is identified as the dominant
intermediate resonance, where the $a_0(980)^+$ contribution is found to be significantly larger than that of the $a_0(980)^0$ state, i.e. $\mathcal{B}(D^+\to a_0(980)^+\pi )/\mathcal{B}(D^+\to a_0(980)^0\pi^+)=2.6\pm0.6\pm 0.3$~\cite{BESIII:2024tpv}, exhibiting a notable discrepancy with the earlier theoretical predictions~\cite{Ikeno:2021kzf}. 

On the other hand, the invariant mass distributions of $\pi^0\eta $ and $\pi^+\eta$ measured by BESIII display a pronounced  enhancement structure around 1.6~GeV~\cite{BESIII:2024tpv}. Considering the $a_0(1710)$ was first observed in the $\eta\pi$ final states by BaBar, and it mass and width are $1704\pm 5 \pm 2$~MeV and $110 \pm 15 \pm 11$~MeV, respectively~\cite{BaBar:2021fkz}, it is expected that the resonance $a_0(1710)$ could contribute to this enhancement structure. Additionally, the $\pi^+\pi^0$ invariant mass distribution deviates from the phase space distribution in the region of $0.6\sim0.8$~GeV~\cite{BESIII:2024tpv}, which implies that the intermediate meson $\rho$ should play a role in this process.

 Thus, in this work we will investigate the Cabibbo-suppressed process $D^+\to \pi^0\pi^+\eta$ by taking into account the $a_0(980)$, dynamically generated from the $S$-wave pseudoscalar meson-pseudoscalar meson interaction with the chiral unitary approach, and the contributions from the intermediate resonances $a_0(1710)$ and $\rho$. 
 
This paper is organized as follows. In Sec.~\ref{sec2}, we will present the mechanism for the process $D^+\to \pi^0\pi^+\eta$, and our results and discussions will be shown in Sec.~\ref{sec.3}, followed by a short summary in the last section.



	\section{Formalism}\label{sec2}
 \subsection{Mechanism of $D^+ \to a_0(980)^{0(+)} \pi^{+(0)} \to \pi^0\pi^+\eta $}
 \label{sec2a}	
	\begin{figure*}[htbp]
		\centering
		\subfigure[]{\includegraphics[scale=0.48]{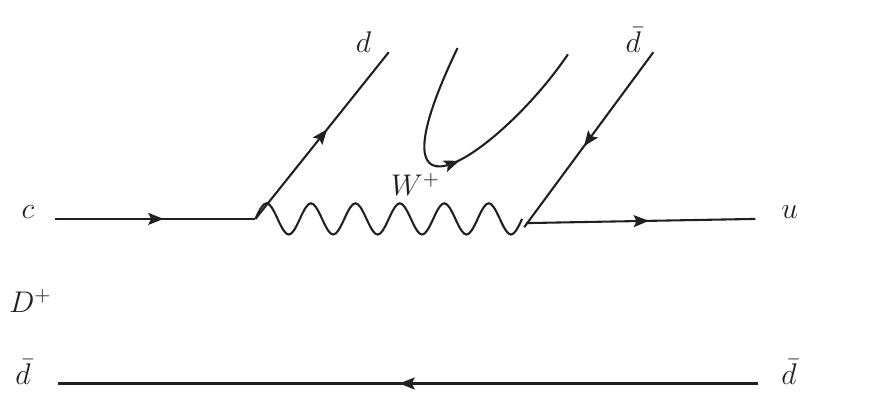}\label{fig:2a}}
		\subfigure[]{\includegraphics[scale=0.48]{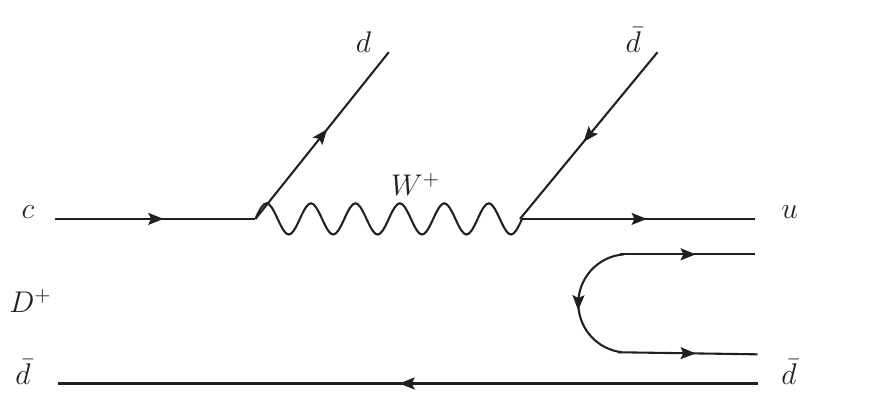}\label{fig:2b}}
		\subfigure[]{\includegraphics[scale=0.48]{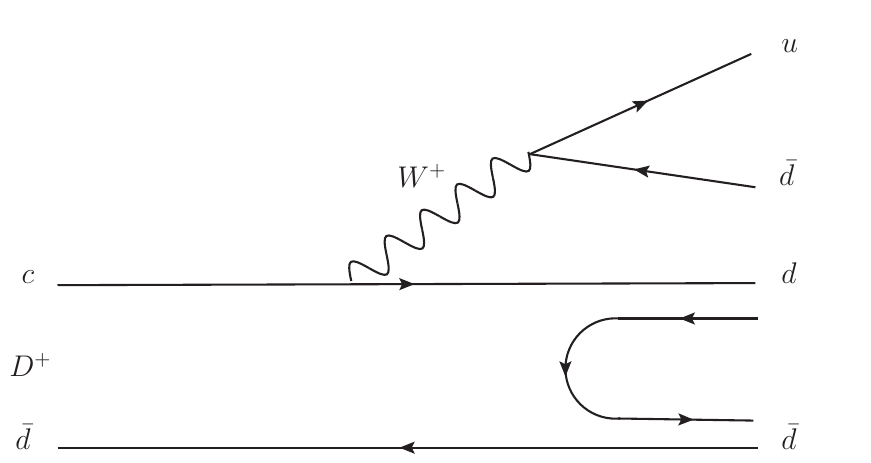}\label{fig:2c}}
		\subfigure[]{\includegraphics[scale=0.48]{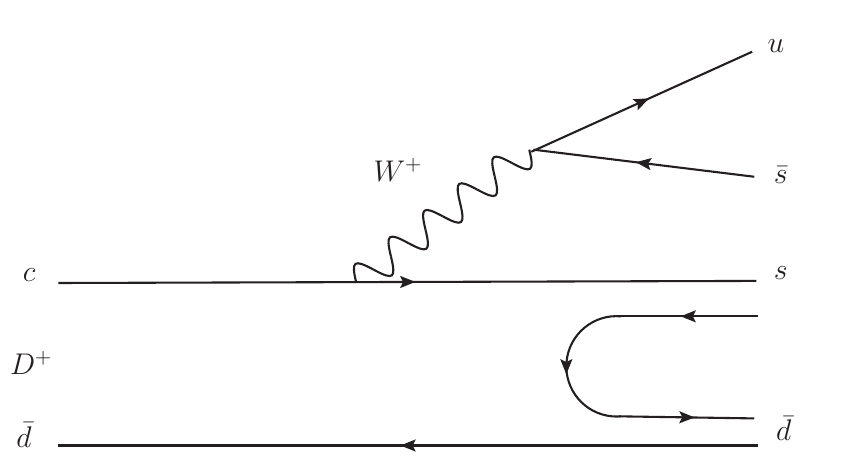}\label{fig:2d}}
		\caption{\small{Diagrammatic representation of the $D^+$ decay. (a) The internal emission of $D^+\to\pi^+d\bar{d}$ and hadronization of the $d\bar{d}$ through $\bar{q}q$ with vacuum quantum numbers; (b) The internal emission of $D^+\to\pi^0u\bar{d}$ and hadronization of the $u\bar{d}$ through $\bar{q}q$ with vacuum quantum numbers; (c) The external emission of $D^+\to\pi^+d\bar{d}$ and hadronization of the $d\bar{d}$ through $\bar{q}q$ with vacuum quantum numbers; (d) The external emission of $D^+\to K^+s\bar{d}$ and hadronization of the $s\bar{d}$ through $\bar{q}q$ with vacuum quantum numbers. }}
		\label{fig:2}
	\end{figure*}
 
In this section, we will demonstrate the decay mechanism of the single Cabibbo-suppressed process $D^+\to \pi^0\pi^+\eta$, which can be divided into three steps, the weakly decay, the hadronization, and the final state interactions~\cite{Duan:2020vye,Ikeno:2021kzf,Molina:2019udw,Wang:2021naf,Zhang:2022xpf}.
For the first step, the $c$ quark in the $D^+$ meson weakly decays into a $d(s)$ quark and a $W^+$ boson, and then the $W^+$ goes to $\bar{d}(\bar{s})$ and $u$ quarks. In order to generate the corresponding final states, all quarks, along with quark pairs $q\bar{q}$ created from the vacuum with the quantum numbers $J^{PC}=0^{++}$, are hadronized into hadrons, which can be classified as the $W^+$ internal emission of Figs.~\ref{fig:2a}-\ref{fig:2b}, and $W^+$ external emission of Figs.~\ref{fig:2c}-\ref{fig:2d}.

	For the internal emission depicted in Fig.~\ref{fig:2a} and Fig.~\ref{fig:2b}, the $u\bar{d}$ and $d\bar{d}$ quarks hadronize into $\pi^+$ and $\pi^0$, while hadronization of other quarks can be written as,
	\begin{equation}\label{eq1}
		\begin{aligned}
		\sum_{i}d(\bar{q}_iq_i)\bar{d}=\sum_{i}M_{2i}M_{i2}=(M^2)_{22},
		\end{aligned}
	\end{equation}
	\begin{equation}\label{eq2}
	\begin{aligned}
		\sum_{i}u(\bar{q}_iq_i)\bar{d}=\sum_{i}M_{1i}M_{i2}=(M^2)_{12},	
	\end{aligned}
	\end{equation}
	where $i=1$, $2$, $3$ represent $u$, $d$, $s$ quarks, respectively. The matrix $M$ can be expressed with the pseudoscalar meson as follows~\cite{Dai:2021owu},	
	\begin{eqnarray}\label{eq3}
		M=
		\begin{pmatrix}
			\dfrac{\pi^0}{\sqrt{2}}+\dfrac{\eta}{\sqrt{3}}+\dfrac{\eta'}{\sqrt{6}}
			&\pi^+  & K^+\\
			\pi^-   &-\dfrac{\pi^0}{\sqrt{2}}+\dfrac{\eta}{\sqrt{3}}+\dfrac{\eta'}{\sqrt{6}}  & K^0\\
			K^- & \bar{K}^0 & -\dfrac{\eta}{\sqrt{3}}+\dfrac{2\eta'}{\sqrt{6}}  \nonumber\\
		\end{pmatrix} .\\
	\end{eqnarray}
	Because the $\eta'$ has a large mass and do not play a significant role in the generation of $a_0(980)$~\cite{Oller:1997ti}, the components of $\eta'$ are omitted in this work. So the Eq.~(\ref{eq1}) and Eq.~(\ref{eq2}) can be re-written as,
	\begin{equation}\label{eq4}
     (M^2)_{22}=\pi^+\pi^-+\dfrac{1}{2}\pi^0\pi^0-\sqrt{\dfrac{2}{3}}\pi^0\eta+\dfrac{1}{3}\eta\eta+K^0\bar{K}^0,
   	\end{equation}	   	
	\begin{equation}\label{eq5}
	(M^2)_{12}=\dfrac{2}{\sqrt{3}}\pi^+\eta+K^+\bar{K}^0	.
	\end{equation}	
	Because the $\pi^+\pi^-$, $\pi^0\pi^0$, and $\eta\eta$ channels in Eq.~(\ref{eq4}) cannot be coupled to the isospin $I=1$ system and do not contribute to the production of $a_0(980)$, we do not consider these channels. Therefore, the possible hadron compositions are as follows,
	\begin{equation}\label{eq.Ha}
	H^{(a)}=V^{(a)}V_{cd}V_{ud}\left(-\sqrt{\dfrac{2}{3}}\pi^0\eta+K^0\bar{K}^0\right)\pi^+,
	\end{equation}	
	\begin{eqnarray}\label{eq.Hb}
H^{(b)}&=&V^{(b)}V_{cd}V_{ud}\left(\dfrac{2}{\sqrt{3}}\pi^+\eta+K^+\bar{K}^0\right)\left(-\dfrac{1}{\sqrt{2}}\pi^0\right)\nonumber\\
&=&V^{(b)}V_{cd}V_{ud}\left(-\sqrt{\dfrac{2}{3}}\pi^+\eta-\dfrac{1}{\sqrt{2}}K^+\bar{K}^0\right)\pi^0		.\end{eqnarray}
The dynamic vertex factors $V^{(a)}$ and $V^{(b)}$ in Eq.~(\ref{eq.Ha}) and Eq.~(\ref{eq.Hb}) correspond to the contributions illustrated in Fig.~\ref{fig:2a} and Fig.~\ref{fig:2b}, respectively, encompassing all relevant factors. 
It is notable that $V^{(a)}$ and $V^{(b)}$ are expected to exhibit similarities, given that their weak decay processes are the same. 
The elements of the CKM matrix are $V_{cd} = V_{us}$ and $V_{ud}=V_{cs}$ up to the order of $\mathcal{O}(\lambda^3)$~\cite{ParticleDataGroup:2024cfk}. 
	
In addition to the mechanisms of the $W^+$ internal emission, the ones of the $W^+$ external emission also contribute to the formation of the $a_0(980)$ in the process $D^+\to \pi^+\pi^0\eta$. For the mechanisms of the $W^+$ external emission in Figs.~\ref{fig:2c} and \ref{fig:2d}, the hadronization step of the Fig.~\ref{fig:2c} is the same as the one of Fig.~\ref{fig:2a}. However, external emission is color-favored relative to internal emission, so the color factor $C$ is introduced to represent the weight of external emission relative to internal emission. For the $W^+$ external emission, the quark pairs $u\bar{d}$ derived from the decay of $W^+$ can be sufficient for the singlet state of $\pi^+$, where $u$ and $\bar{d}$ can have three choices: red, green, and blue. However, the $u$ and $\bar{d}$ from the $W^+$ internal emission have fixed colors, and we take $C=3$ in case of color number $N_c=3$~\cite{Wei:2021usz,Dai:2018nmw,Zhang:2020rqr,Duan:2020vye}.
For Fig.~\ref{fig:2d}, except for the different elements of the CKM matrix, everything else is the same as Fig.\ref{fig:2c}. So for the $W^+$ external emission process in Fig.~\ref{fig:2c} and Fig.~\ref{fig:2d}, the final compositions are as follows,
	\begin{eqnarray}\label{eq.Hc}
		H^{(c)}=C\times V^{(a)}V_{cd}V_{ud}\left(-\sqrt{\dfrac{2}{3}}\pi^+\pi^0\eta+K^0\bar{K}^0\pi^+\right),
	\end{eqnarray}	
	\begin{eqnarray}\label{eq.Hd}
   H^{(d)}&=&C\times V^{(a)}V_{cs}V_{us}\left(M^2\right)_{32}K^+\nonumber\\
       &=&C\times V^{(a)}V_{cs}V_{us}\left(K^+K^-\pi^+-\dfrac{1}{\sqrt{2}}K^+\bar{K}^0\pi^0\right).\nonumber\\
	\end{eqnarray}
	Then we have all the possible components after the hadronization, 	
	\begin{eqnarray}\label{eq:totalH}
	H&=&H^{(a)}+H^{(b)}+H^{(c)}+H^{(d)}\nonumber\\
	&=&V^{(a)}\left[-\sqrt{\dfrac{2}{3}}(1+C+R)\pi^+\pi^0\eta+(1+C)K^0\bar{K}^0\pi^+ \right.\nonumber\\
	&& \left. +C\times K^+K^-\pi^+-\dfrac{1}{\sqrt{2}}(C+R)K^+\bar{K}^0\pi^0\right].
	\end{eqnarray}
	
	\begin{figure}[tbhp]
		\centering
		\subfigure[]{\includegraphics[scale=0.47]{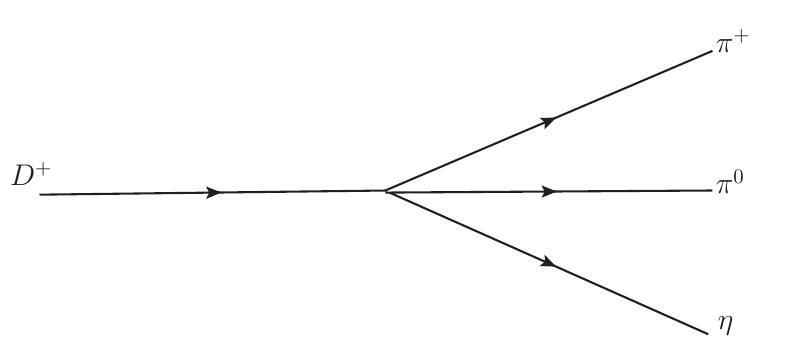}}
		\subfigure[]{\includegraphics[scale=0.47]{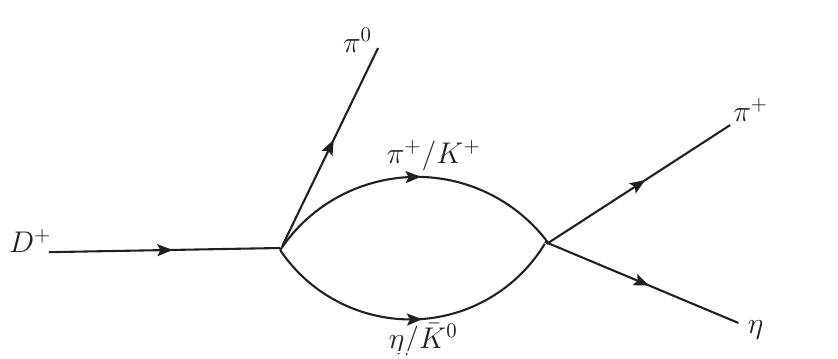}}
		\subfigure[]{\includegraphics[scale=0.47]{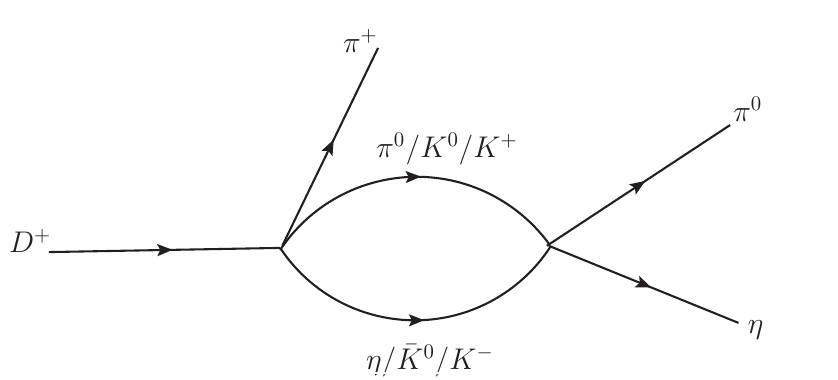}}
		\caption{\small{Mechanisms of the $D^+\to\pi^+\pi^0\eta$: (a) tree diagram, (b) the final state interaction of $\pi^+\eta$, $K^+\bar{K}^0$, and (c) the final state interaction of $\pi^0\eta$, $K^0\bar{K}^0$, and $K^+K^-$.}}
		\label{fig:3}
	\end{figure}

	The elements of the CKM matrix are shrunk into a normalization factor of $V^{(a)}$ in Eq.~(\ref{eq:totalH}), where the $R={V_{(b)}}/{V_{(a)}}$ represents the relative weight of Fig.~\ref{fig:2b} to Fig.~\ref{fig:2a}, and $C$ is the color factor mentioned above. The weak processes of Fig.~\ref{fig:2a} and Fig.~\ref{fig:2b} are the same, so in this work, we take $R={V_{(b)}}/{V_{(a)}}=1$. 
From Eq.~(\ref{eq:totalH}), one can find the final states of the process $D^+\to \pi^0\pi^+\eta$ could be produced directly with the relative strength $-\sqrt{2/3}(1+C+R)$. As depicted in Fig.~\ref{fig:3}(a), we can write the amplitude for the tree diagram as,
\begin{equation}
    \mathcal{T}_{\rm tree}=-\sqrt{\dfrac{2}{3}}(1+C+R)V^{(a)}.
\end{equation}
As depicted in Figs.~\ref{fig:3}(b) and \ref{fig:3}(c), we can write the transition amplitude of generating the $a_0 (980)^{+,0}$ in the process $D^+\to \pi^0\pi^+\eta$ as follows, 
	\begin{eqnarray}\label{(980)0amp}
	&\mathcal{T}_{a_0(980)^0}=V^{(a)}\left[h_{\pi^0\pi^+\eta}G_{\pi^0\eta}(M_{\pi^0\eta})t_{\pi^0\eta\to\pi^0\eta}(M_{\pi^0\eta})\right.\nonumber\\
&+h_{K^0\bar{K}^0\pi^+}G_{K^0\bar{K}^0}(M_{\pi^0\eta})t_{K^0\bar{K}^0\to\pi^0\eta}(M_{\pi^0\eta})\nonumber\\
&\left. +h_{K^+K^-\pi^+}G_{K^+K^-}(M_{\pi^0\eta})t_{K^+K^-\to\pi^0\eta}(M_{\pi^0\eta})\right],
	\end{eqnarray}	
	\begin{eqnarray}\label{(980)+amp}
	&\mathcal{T}_{a_0(980)^+}=	V^{(a)}\left[h_{\pi^0\pi^+\eta}G_{\pi^+\eta}(M_{\pi^+\eta})t_{\pi^+\eta\to\pi^+\eta}(M_{\pi^+\eta})\right.\nonumber\\
	&\left.+h_{K^+\bar{K}^0\pi^0}G_{K^+\bar{K}^0}(M_{\pi^+\eta})t_{K^+\bar{K}^0\to\pi^+\eta}(M_{\pi^+\eta})\right],		
	\end{eqnarray}	
where $h_{\pi^0\pi^+\eta}= -\sqrt{\frac{2}{3}} ( 1+C+R )$, $h_{K^0\bar{K}^0\pi^+}=1+C$, $h_{ K^+K^-\pi^+}=C $, and $h_{ K^+\bar{K}^0\pi^0}=-\frac{1}{\sqrt{2}}(C+R)$, obtained from Eq.~(\ref{(980)0amp}) and Eq.~(\ref{(980)+amp}). The $G_i$ is the loop function of two-meson propagator, and $t_{i\to j}$ represents the transition amplitude from channel $i$ to channel $j$. The $t$-matrix can be obtained by solving the Bethe-Salpeter equation in coupled channels~\cite{Oset:2016lyh},
\begin{eqnarray}\label{T}
	T=[1-VG]^{-1}V,
\end{eqnarray}
where the $V$ is  a $2\times 2$ matrix of interaction kernel potential between isospin channels $K\bar{K}$ and $\pi\eta$~\cite{Oller:1997ti}. Because of isospin multiple states $K=(K^+,K^0)$,
$\bar{K}=(\bar{K}^0,-K^-)$, and $\pi=(-\pi^+,\pi^0,\pi^-)$, this matrix $V$ can also be expressed as~\cite{Xie:2014tma},
\begin{equation}
	\begin{aligned}
		&V_{K\bar{K}\to K\bar{K}}=-\dfrac{1}{4f^2}s,\\
		&V_{K\bar{K}\to \pi\eta}=\dfrac{\sqrt{6}}{12f^2}\left(3s-\dfrac{3}{8}m^2_K-\dfrac{1}{3}M^2_{\pi}-m^2_{\eta}\right),\\
		&V_{\pi\eta\to \pi\eta}=-\dfrac{1}{3f^2}m^2_{\pi},
	\end{aligned}
\end{equation}
with the pion decay constant $f=93$~MeV and the invariant mass square $s$ of the meson-meson system.
There is also a conversion method between charge channels and isospin channels,
	\begin{eqnarray}
	\begin{aligned}
		&t_{K^+K^-\to \pi^0\eta}=-\dfrac{1}{\sqrt{2}}t^{I=1}_{K\bar{K}\to\pi\eta},\\
		&t_{K^0\bar{K}^0\to\pi^0\eta}=\dfrac{1}{\sqrt{2}}t^{I=1}_{K\bar{K}\to\pi\eta},\\
		&t_{K^+\bar{K}^0\to\pi^+\eta}=-t^{I=1}_{K\bar{K}\to\pi\eta}. 
	\end{aligned}
   \end{eqnarray}
     
 The loop function $G_i$ for two-mesons in Eq.~(\ref{T}) is given by,
 \begin{equation}
	G_i=i\int \dfrac{d^4q}{(2\pi^4)}\dfrac{1}{(q-P)^2-m^2_2+i\epsilon}\dfrac{1}{q^2-m^2_1+i\epsilon},
\end{equation}
where the $m_1$, $m_2$ are the masses of mesons in the $i$-channel, $q$ is the four-momentum of one meson, and the $P$ is the total four-momentum. Since the loop function $G_i$ in Eq.~(\ref{(980)0amp}) and Eq.~(\ref{T}) is logarithmically divergent, we have two methods to solve this singular integral, either using the three-momentum cut-off method, or the dimensional regularization method~\cite{Dias:2016gou,Ahmed:2020kmp,Oller:2000fj}. The choice of method does not significantly impact our results. In this work, we use the method of dimensional regularization method, where the loop function $G_i$ can be written as,   
    \begin{eqnarray}
    \begin{aligned}
 G_i&=\dfrac{1}{16\pi^2}\{a_i+\ln\dfrac{m^2_1}{\mu^2}+\dfrac{m^2_2-m^2_1+s}{2s}\ln\dfrac{m^2_2}{m^2_1}\\
 	&+\dfrac{p}{\sqrt{s}}[\ln(s-(m^2_2-m^2_1)+2p\sqrt{s})\\
 	&+\ln (s+(m^2_2-m^2_1)+2p\sqrt{s})\\
 	&-\ln (-s+(m^2_2-m^2_1)+2p\sqrt{s})\\
 	&-\ln(-s-(m^2_2-m^2_1)+2p\sqrt{s})] \}
    \end{aligned}
    \end{eqnarray}
    where $p={\sqrt{(s-(m_1+m_2)^2)(s-(m_1-m_2)^2)}}/{2\sqrt{s}}$, and $\mu$ is the dimension normalization scale, according to Ref.~\cite{Gamermann:2006nm}, we have $\mu=600$~MeV, $a_{\pi\eta}=-1.71$, and $a_{K\bar{K}}=-1.66$.
    
    According to the previous discussion, the scattering amplitude of $a_0(980)$ can be written as,   
    \begin{eqnarray}\label{eq:(980)0amp2}
    	&\mathcal{T}_{a_0(980)^0}=V^{(a)}\left[ h_{\pi^0\pi^+\eta} G_{\pi^0\eta}(M_{\pi^0\eta})t^{I=1}_{\pi\eta\to \pi\eta }(M_{\pi^0\eta}) \right.\nonumber \\
    	& \left.+\frac{h_{K^0\bar{K}^0\pi^+}}{\sqrt{2}}   G_{K^0\bar{K}^0}(M_{\pi^0\eta})t ^{I=1}_{K\bar{K}\to \pi\eta }(M_{\pi^0\eta})\right],
      \end{eqnarray}    
    \begin{eqnarray}\label{eq:(980)+amp2}
    	&\mathcal{T}_{a_0(980)^+}=V^{(a)}\left[ h_{\pi^0\pi^+\eta} G_{\pi^+\eta}(M_{\pi^+\eta})t^{I=1}_{\pi\eta\to \pi\eta }(M_{\pi^+\eta}) \right. \nonumber \\
    &- \left.  h_{ K^+\bar{K}^0\pi^0} G_{K^+\bar{K}^0}(M_{\pi^+\eta})t^{I=1}_{ K\bar{K}\to \pi\eta }(M_{\pi^+\eta}) \right].
    \end{eqnarray}

It is notable that, the most reliable prediction range of the  $S$-wave pseudoscalar meson-pseudoscalar meson interaction within the chiral unitary approach is up to $1100\sim1200$~MeV, and one can not use this model for higher invariant masses~\cite{Debastiani:2016ayp}. 
    In order to present a reasonable $\pi^0\eta$ and $\pi^+\eta$ invariant mass distributions, we evaluate $Gt(M_{\rm inv})$ combinations up to $M_{\rm inv}=M_{\rm cut}$, and from there we multiply $Gt$ by a smooth factor to make it gradually decreasing at large $M_{\pi^0\eta}$ and $M_{\pi^+\eta }$, as done in,
    \begin{equation}
    	Gt(M_{\rm inv})=Gt(M_{\rm cut})e^{-\alpha (M_{\rm inv}-M_{\rm cut})},~~  M_{\rm inv}>M_{\rm cut}, \label{eq:cut}
    \end{equation} 
Here, we take the value $M_{\rm cut}=1150$ MeV and the smoothing factor $\alpha$=0.0054~MeV~\cite{Debastiani:2016ayp}.

\subsection{Mechanism of the intermediate $\rho$ and $a_0(1710)$}
 \label{sec2b}
\begin{figure}[htbp]
 	\centering 	
 		\centering
 	\includegraphics[scale=0.55]{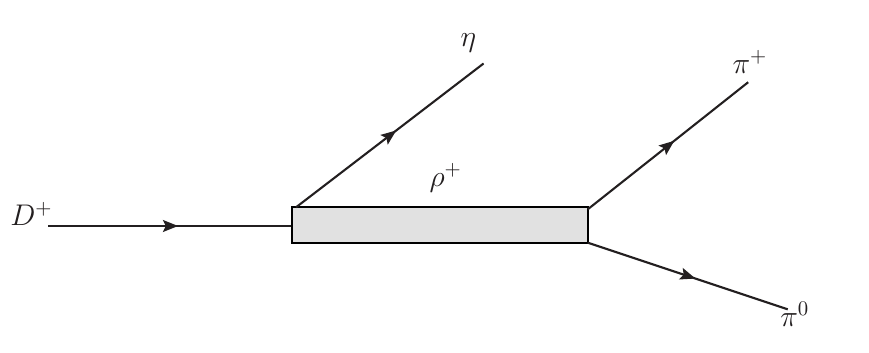}
 	\caption{Process $D^+\to \rho^+ \eta \to \pi^0\pi^+\eta$ via the intermediate vector $\rho^+$.}\label{fig:rho}
 \end{figure}

In the process $D^+\to \pi^0\pi^+\eta$, in addition to considering the contribution of $a_0(980)$, we also take into account the contribution of $\rho^+$ and $a_0(1710)$. The hadron level diagram for the process $D^+\to \rho^+ \eta \to \pi^0\pi^+\eta$ is depicted in Fig.~\ref{fig:rho}, and the decay amplitude can be written as~\cite{Zhang:2024myn},
\begin{eqnarray} \label{eq:rho}
\mathcal{T}^{\rho^+}&=&\dfrac{g_v}{M_{\pi^0\pi^+}^2-m^2_{\rho^+}+iM_{\rho^+}\Gamma_{\rho^+}}\nonumber\\	&&\left[\left(m^2_{\pi^+}-m^2_{\pi^0}\right)\left(1-\dfrac{M^2_{\pi^0\pi^+}}{m^2_{\rho^+}}\right)\right. \nonumber \\
&& + 2p_{\pi^0}\cdot p_{\eta} \dfrac{m^2_{\pi^0}-m^2_{\pi^+}-m^2_{\rho^+}}{m^2_{\rho^+}}\nonumber\\
 &&\left. +2p_{\pi^+}\cdot p_{\eta}\dfrac{m^2_{\pi^0}-m^2_{\pi^+}+m^2_{\rho^+}}{m^2_{\rho^+}}\right],
 	\end{eqnarray} 
where
 \begin{align}
 &p_{\pi^0}\cdot p_{\eta}=\frac{M^2_{\pi^0\eta}-m^2_{\pi^0}-m^2_{\eta}}{2},\nonumber\\
 		&p_{\pi^+}\cdot p_{\eta}=\frac{m^2_{D^+}+m^2_{\pi^0}-M^2_{\pi^0\eta}-M^2_{\pi^0\pi^+}}{2}.
 \end{align}
  The $g_v$ can be regarded as the weight factor of meson $\rho$, containing the coupling constants of the vertices $D^+\to \rho^+\eta$ and $\rho^+\to \pi^+\pi^0$, and  the $M_{\pi^0\pi^+}$ is the invariant mass of the $\pi^0\pi^+$ system. 
 
As discussed in the introduction, the measured  invariant mass spectrum of $\pi^0\eta $ and $\pi^+\eta$ have significant enhancement structure around 1.6~GeV~\cite{BESIII:2024tpv}, and the scalar meson $a_0(1710)$ could play a role in this process. Thus, we will take into account $a_0(1710)$ with the Breit-Wigner form,
\begin{equation}\label{eq:(1710)0amp}
\mathcal{T}_{a_0(1710)^0}=\dfrac{V^{(b)}_1 \times M^2_{a_0(1710)}}{m^2_{\pi^0\eta}- M^2_{a_0(1710)}+i\Gamma_{a_0(1710)}M_{a_0(1710)}},
\end{equation}
\begin{equation}\label{eq:(1710)+amp}
	\mathcal{T}_{a_0(1710)^+}=\dfrac{ V^{(b)}_2\times M^2_{a_0(1710)}}{m^2_{\pi^+\eta}- M^2_{a_0(1710)}+i\Gamma_{a_0(1710)}M_{a_0(1710)}},
\end{equation}
where the $V^{(b)}_1$and $V^{(b)}_2$ represent the weight proportions of $a_0(1710)^0$ and $a_0(1710)^+$, respectively. The width and mass of $a_0(1710)$ are taken as  $\Gamma_{a_0(1710)}=110$~MeV and $M_{a_0(1710)}=1704$~MeV~\cite{BaBar:2021fkz}.  
	
Finally, the total scattering amplitude of process $D^+\to \pi^+\pi^0\eta $ can be written as,
\begin{equation}\label{T^2}
	\begin{aligned}	\left|\mathcal{T}\right|^2&=|\mathcal{T}_{\rm tree}+\mathcal{T}_{a_0(980)^0}+\mathcal{T}_{a_0(980)^+}e^{i\phi}
	+\mathcal{T}_{\rho^+}e^{i\phi '}\\
 &+\mathcal{T}_{a_0(1710)^0}e^{i\phi ''}+\mathcal{T}_{a_0(1710)^+}e^{i\phi'''}|^2,
		\end{aligned}
\end{equation}
 where the $\phi$, $\phi'$, $\phi''$, $\phi'''$ are the phase angles between different contributions.\footnote{Here we have not considered the relative phase angle between the $\mathcal{T}_{\rm tree}$ and $\mathcal{T}_{a_0(980)^0}$ in order to reduce the number of the free parameters. Indeed, we have performed the fit by adding the relative phase angle between two amplitudes, and found that the fitted $\chi^2/d.o.f.=1.69$ is the same as the following result without this phase angle.}
 Then, the invariant mass distribution of the double differential width of the $D^+\to \pi^+\pi^0\eta$ can be expressed as, 
  \begin{eqnarray}
  	\frac{d^2\Gamma}{dM_{\pi^0\eta}dM_{\pi^+\eta}}&=&\frac{1}{(2\pi)^3} \frac{M_{\pi^0\eta}M_{\pi^+\eta}}{8m^3_{D^+}}{|{\cal T}|}^2,\label{eq:width1}
  \end{eqnarray}  
  \begin{eqnarray}
  	\frac{d^2\Gamma}{dM_{\pi^0\eta}dM_{\pi^+\pi^0}}&=&\frac{1}{(2\pi)^3} \frac{M_{\pi^0\eta}M_{\pi^+\pi^0}}{8m^3_{D^+}}{|{\cal T}|}^2.\label{eq:width2}
  \end{eqnarray}
  
 One could obtain the invariant mass distributions  ${d\Gamma}/{dM_{\pi^0\eta}}$, ${d\Gamma}/{dM_{\pi^+\eta}}$ and ${d\Gamma}/{dM_{\pi^+\pi^0}}$ by integrating Eq.~(\ref{eq:width1}) and  Eq.~(\ref{eq:width2}) over each of the invariant mass variables. For a given value of $M_{12}$, the range of $M_{23}$ is determined by~\cite{ParticleDataGroup:2024cfk}, 
 \begin{align}
 	&\left(m_{23}^2\right)_{\min}=\left(E_2^*+E_3^*\right)^2-\left(\sqrt{E_2^{* 2}-m_2^2}+\sqrt{E_3^{* 2}-m_3^2}\right)^2 ,\nonumber\\
 	&\left(m_{23}^2\right)_{\max}=\left(E_2^*+E_3^*\right)^2-\left(\sqrt{E_2^{* 2}-m_2^2}-\sqrt{E_3^{* 2}-m_3^2}\right)^2,
 \end{align} 
 where $E_2^{*}$ and $E_3^{*}$ are the energies of particles 2 and 3 in the $M_{12}$ rest frame, which are written as,
 \begin{align}
 	&E_2^{*}=\dfrac{M_{12}^2-m_1^2+m_2^2}{2M_{12}}, \nonumber\\
 	&E_3^{*}=\dfrac{M_{D^+}^2-M_{12}^2-m_3^2}{2M_{12}},
 \end{align}
 with $m_1$, $m_2$, and $m_3$ are the masses of involved particles 1, 2 and 3, respectively. All the masses and widths of the particles involved in this work are taken from the Review of Particle Physics~\cite{ParticleDataGroup:2024cfk}.
 
 \section{RESULTS AND DISCUSSIONS}\label{sec.3}
 
\begin{figure*}[htbp] 
 	\subfigure[]{
 		\includegraphics[scale=0.45]{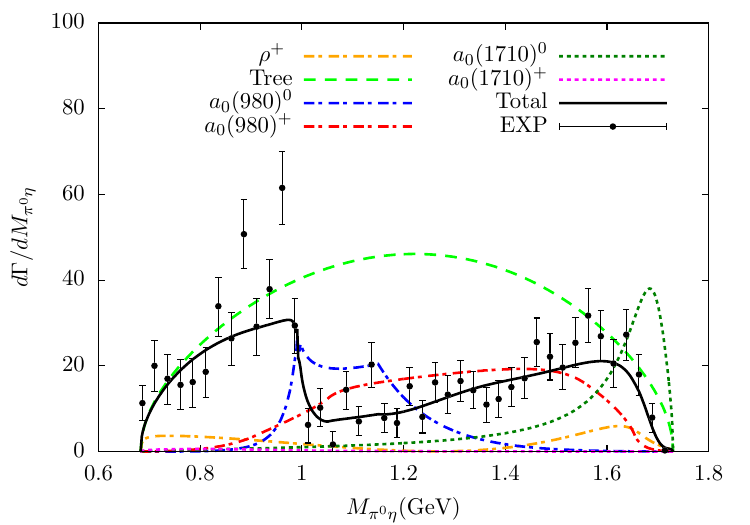}\label{pi0eta(1710)}}
 	\subfigure[]{
 		\includegraphics[scale=0.45]{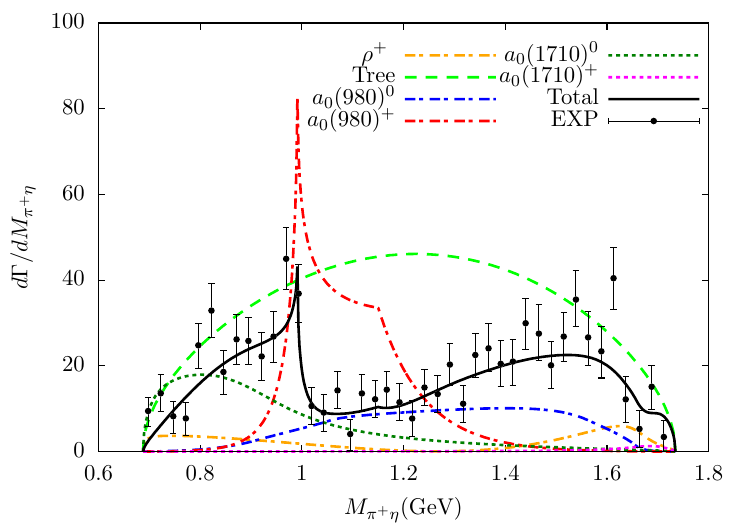}\label{pi+eta(1710)}}
 	\subfigure[]{
 		\includegraphics[scale=0.45]{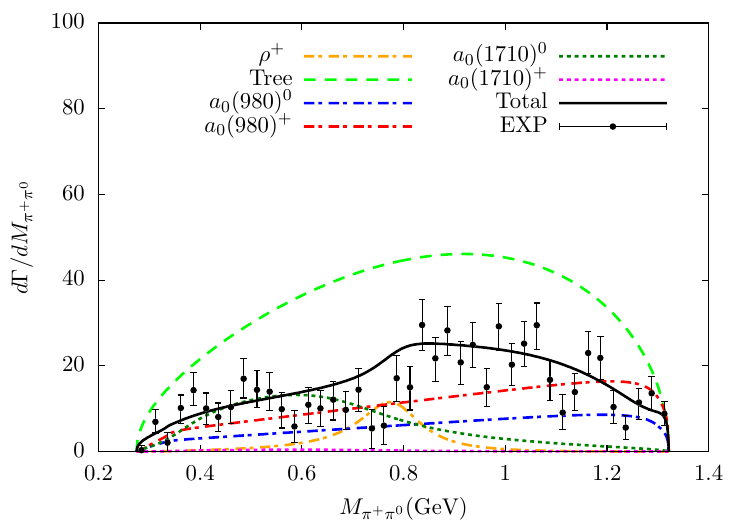}\label{pi0pi+(1710)}}
  \subfigure[]{
  \includegraphics[scale=0.45]{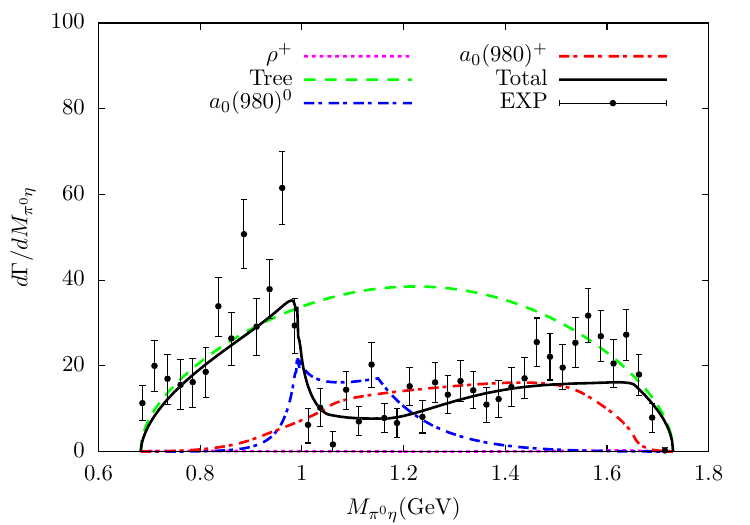}\label{pi0eta(rho)}}
 	\subfigure[]{
 		\includegraphics[scale=0.45]{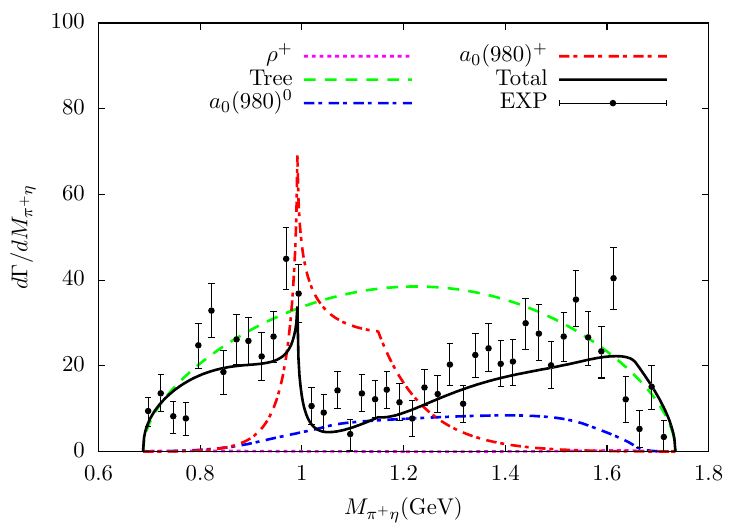}\label{pi+eta(rho)}}
 	\subfigure[]{
  \includegraphics[scale=0.45]{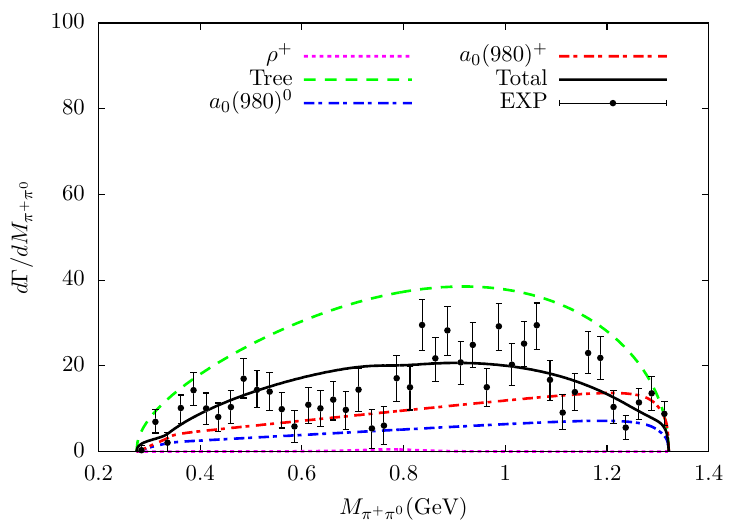}\label{pi0pi(rho)}}
 	\caption{$\pi^0\eta$ (a), $\pi^+\eta$ (b) and $\pi^0\pi^+$ (c) invariant mass distributions for the process $D^+\to \pi^0\eta\pi^+$ considered the $a_0(980)$, $a_0(1710)$, $\rho^+$; $\pi^0\eta$ (d), $\pi^+\eta$ (e) and $\pi^0\pi^+$ (f) invariant mass distributions for the process $D^+\to \pi^0\eta\pi^+$ by turning off the contribution of $a_0(1710)$. The experimental data of BESIII is represented by the points with error bars~\cite{BESIII:2024tpv}. }
 	\label{fig.jieguo1}
 \end{figure*}

\begin{table}[h]
\centering
\caption{Values of the fitting parameters.}\label{tab:par}
\begin{tabular}{c|c|c}
\hline\hline
\multirow{2}{*}{parameter} & \multicolumn{2}{c}{Fit numerical results} \\
\cline{2-3}
 & A & B \\
\hline
$\chi^2 / N_{\text{dof}}$ & $1.69$ & $2.55$ \\
\hline
$V^{(a)}$ & $196.27 \pm 5.73$ & $179.39 \pm 2.55$ \\
$V^{(b)}_1$ & $84.66 \pm 8.31$ & -- \\
$V^{(b)}_2$ & $15.33 \pm 13.68$ & --\\
$g_{v}$ & $36.04 \pm 6.66$ & $7.81 \pm 3.78$ \\
$\phi$  & $(0.29 \pm 0.02)\pi$ & $(0.40 \pm 0.03)\pi$ \\
$\phi'$ & $(-1.14 \pm 0.10)\pi$ & $(0.31 \pm 0.14)\pi$ \\
$\phi''$ & $(-0.42 \pm 0.50)\pi$ & -- \\
$\phi'''$ & $(-1.08 \pm 0.15)\pi$ & -- \\  
\hline \hline
\end{tabular}
\label{table1}
\end{table}

	 \begin{figure*}[htbp]
     \centering
     \subfigure[]{
	\includegraphics[scale=0.9]{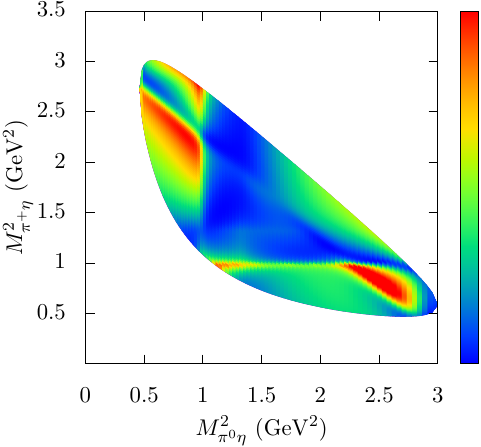}\label{Dalitz(1710)}}
 \subfigure[]{
 \includegraphics[scale=0.9]{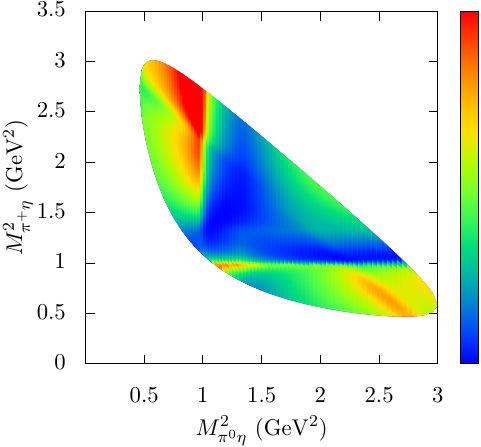}\label{Dalitz(rho)}}
 	\caption{Dalitz plots of `$M^2_{\pi^0\eta}$' vs. `$M^2_{\pi^+\eta}$' for the process $D^+\to \pi^0\pi^+\eta$. The results of Fig.~\ref{Dalitz(1710)} are obtained from our full model, and Fig.~\ref{Dalitz(rho)} are obtained by turning off the contribution of the $a_0(1710)$.}
  \label{fig:dalitz}
 \end{figure*}
 In this work, there are eight parameters, (1) $V^{(a)}$, the factor normalization of $a_0(980)$ in Eq.~(\ref{eq:(980)0amp2}) and Eq.~(\ref{eq:(980)+amp2}); (2) $g_v$ represents the weight of $\rho^+$ in Eq.~(\ref{eq:rho});   (3) $V^{(b)}_1$and $V^{(b)}_2$ represent the weights of the contributions from $a_0(1710)^0$ and $a_0(1710)^+$ in Eq.~(\ref{eq:(1710)0amp}) and Eq.~(\ref{eq:(1710)+amp}), respectively;
 (4) four phase angles $\phi$, $\phi'$, $\phi''$, $\phi'''$ for the interference between different contributions in Eq.~(\ref{T^2}). 
 To demonstrate our theoretical results, we have fitted the $\pi^0\eta$, $\pi^+\eta$, $\pi^+\pi^0$ invariant mass distributions measured by BESIII, where the background are subtracted. Then we obtained $\chi^2/d.o.f. =200.53/(42+42+43-9)=1.69$, and the fitted parameters are shown in Table~\ref{tab:par} (Case A). 

 Then we have calculated the $\pi^0\eta$, $\pi^+\eta $, and $\pi^+\pi^0$ invariant mass distributions with the fitted parameters, as present in Figs.~\ref{fig.jieguo1}(a-c). The black points with error bars labeled as ``BESIII data'' are the BESIII data taken from Ref.~\cite{BESIII:2024tpv}, and the black solid line labeled ``Total" represents the results of total contributions. One can find that, in the $\pi^0\eta$ (Fig.~\ref{pi0eta(1710)}) and $\pi^+\eta$ (Fig.~\ref{pi+eta(1710)}) invariant mass distributions, our results show a significant peak structure around 980~MeV, which can be associated with $a_0(980)^0$ and $a_0(980)^+$. Our results of the $\pi^+\pi^0$ invariant mass distribution are in reasonable agreement with the BESIII measurements. In addition, one can find that  the scalar $a_0(1710)^0$ plays a significant role for the enhancement structure in the $\pi^0\eta$ invariant mass distribution around 1.6~GeV.

On the other hand, we have performed another fit by turning off the contribution from the $a_0(1710)$, and present the fitted parameters in Table~\ref{tab:par} (Case B). 
In this case, the fitted $\chi^2/d.o.f. =313.84/(42+42+43-4)=2.55$, which is larger than the previous result. The $\pi^0\eta$, $\pi^+\eta $, and $\pi^+\pi^0$ invariant mass distributions with the fitted parameters of Case B are shown in Figs.~\ref{fig.jieguo1}(d-f). One can find that, although the peaks of the $a_0(980)$ in the $\pi^0\eta$ and $\pi^+\eta$ invariant mass distribution can be well described, the enhancement structure around 1.6~GeV can not be well reproduced. Thus, one can conclude that the $a_0(1710)$ should play an important role in this process.

It is notable that the $\eta\pi$ scattering amplitude is largely unknown at the higher mass tail of the $\eta\pi$ mass distributions, and we introduce an exponential dampening factor to make it gradually decreasing at higher  $\eta\pi$ mass distributions, as shown in Eq.~(\ref{eq:cut}). Thus, the smoothing factor $\alpha$ could give rise to the uncertainties in the analysis. Then we make the fitting by taking different values of $\alpha=0.0037,0.0054,0.0077$, and obtain the fitted $\chi^2 / d.o.f.= 1.68, 1.69, 1.78$,  respectively, which implies that the uncertainties from the smoothing factor $\alpha$ should be very small.

Furthermore, we have also predicted the Dalitz plots of the process $D^+\to \pi^0 \eta \pi^+ $ with and without the contribution of $a_0(1710)$ in Figs.~\ref{fig:dalitz}(a) and \ref{fig:dalitz}(b), respectively. One can find the clear signals of $a_0(980)$ in both figures. However, one can find that the $a_0(1710)^0$ gives an important contribution around $M^2_{\pi^0\eta}=2.7$~GeV$^2$ in Fig.~\ref{fig:dalitz}(a).

With the fitted parameters of Case A, we have calculated the the ratio $\mathcal{B}(D^+\to a_0(980)^+\pi )/\mathcal{B}(D^+\to a_0(980)^0\pi^+)=1.87$, which is consistent with the BESIII data $\mathcal{B}(D^+\to a_0(980)^+\pi )/\mathcal{B}(D^+\to a_0(980)^0\pi^+)=2.6\pm0.6\pm 0.3$ within the uncertainties. Thus, the large $a_0(980)^+$ contribution with respect to the $a_0(980)^0$ could be explained within our theoretical model.

Although our results are in agreement the BESIII measurements, it is notable that the BESIII data have large experimental errors and large fluctuation. Considering that the Belle/Belle II and LHCb have more data sample,  the more precise measurements of this process could shed light on the scalar mesons $a_0(980)$ and $a_0(1710)$ production in this reaction.

\section{Summary And Conclusions }\label{sec.4}
 Motivated by the BESIII measurements about the process $D^+\to \pi^0\pi^+\eta$, we have investigated the single Cabibbo-suppressed process  by taking into account  the final state interactions of the meson-meson interaction in couple channels with the chiral unitary approach, which will dynamically generates the scalar $a_0(980)$. In addition, we also consider the contributions of intermediate resonance $\rho$ and $a_0(1710)$. 
 
 Based on the decay mechanism of this process $D^+\to \pi^0\pi^+\eta$, we calculated the invariant mass distributions of $\pi^0\eta$, $\pi^+\eta$, and $\pi^+\pi^0$, and found a clear peak near 980 MeV, which is associated with scalars $a_0(980)^0$ or $a_0(980)^+$.
Meanwhile, one can conclude that  the scalar $a_0(1710)^0$ plays an significant role for the enhancement structure in the $\pi^0\eta$ invariant mass distribution around 1.6~GeV.
 More precise experimental measurements of this process by Belle~II~\cite{Belle-II:2018jsg} or proposed STCF~\cite{Lyu:2021tlb} could provide deeper insight into the nature of the scalar mesons $a_0(980)$ and $a_0(1710)$.


	\section*{Acknowledgements}
	
This work is supported by the National Key R\&D Program of China (No. 2024YFE0105200),  the Natural Science Foundation of Henan under Grant No. 232300421140, the National Natural Science Foundation of China under Grant No. 12475086 and No. 12192263.

\end{document}